\documentclass{article}

\begin{document}
\title{Exact plane symmetric cosmological solutions for a massless scalar field and cosmological constant}

\author{Chris Vuille, Ph.D.\\ Department of Physical Sciences\\Embry-Riddle Aeronautical
 University,\\Daytona Beach, FL 32114\\vuille@erau.edu \and
Jocelyn Dunn\\Department of Aerospace Engineering\\Embry-Riddle Aeronautical University
\\Daytona Beach, FL 32114\\Joceydunn@yahoo.com}
\maketitle

\abstract{The Klein-Gordon equations were recently solved in general relativity for the case of a plane-symmetric
  static massless scalar field with cosmological constant. By analytic continuation, time-dependent solutions
can be obtained that correspond to cosmologies.  These spacetimes are studied here, and include among them a
 solution exhibiting both early inflation and later continued rapid expansion that matches well with the Concordance Model.}

\section{Introduction}

In recent years observers have used distant supernovae to accurately measure the rate of expansion of the universe. \cite{Riess}
It has been found that the universe is expanding rapidly now, not just in the very early universe.  The unknown
 material that drives that expansion is called dark energy, and is thought to comprise nearly three-quarters
 of the matter in the known universe.  Dark energy has become one of the most important areas of research in science
 today. Further, models developed in efforts to understand it are mainly empirical: there is as yet no quantum theory
 of dark energy. 
   
Exact solutions of Einstein's Equations have long been a guide to the visualization and understanding of the role
of gravitation in the universe.  Much of this work has been collected in Stefani et al. \cite{Exact}. Plane-symmetric space-times
 have been studied by numerous authors.
 Taub \cite{Taub}, in 1951, found the general vacuum solution. A generalization to a spacetime with 
cosmological constant was found by Novotny and Horsky \cite{Novotny} in 1974, and the solution for
 a plane-symmetric static field was found by Singh \cite{Singh} in the same year. Recently Vuille (\cite{Vuille} found
the generalization of all these space-times, and in addition showed that cosmological solutions could be 
generated from them.  All five of the cosmological solutions will be presented here and their salient properties
deduced. Two of the solutions are essentially identical.  One such solution exhibits both rapid inflation in the early universe
and the effects of dark energy.

\section{Cosmologies from the First Solution}
The details of the plane symmetric spacetime of a massless scalar field and cosmological constant can be
found in \cite{Vuille}, with conventions on curvature following Carmeli \cite{Carmeli}. The derivation begins with the
metric 
\begin{equation}
ds^2 = e^{\nu} dt^2 -e^{\lambda} dz^2 - Y^2 \left(dx^2 +
dy^2 \right)
\end{equation}
whereas the equation of the massless scalar field is
\begin{equation} \label{E1}
\nabla^a \nabla_a \psi = 0
\end{equation}
Here $\nu$,$\lambda$, $Y$, and $\psi$ are functions of \textit{z}. With various coordinate transformations, it is possible to put
the metric in the form
\begin{equation}
ds^2 = e^\nu dt^2 - e^\nu W^{-4} du^2 - W^{-2}\left(dx^2 + dy^2 \right)
\end{equation}
where $u = \psi(z)$ and $\nu$ and $W$ are both unknown functions of \textit{u}.  The scalar field equation is therefore automatically solved
in these coordinates and there remains only the metric functions to derive with the gravitational field equations.  Following
\cite{Vuille}, those field equations yield the following relationship between $\nu$ and $W$:
\begin{equation}\label{N3}
W = e^{{-\nu/2} +\frac{1}{2}c_1 u + \frac{1}{2} c_2}
\end{equation}
The constant $c_2$ can be eliminated by trivial coordinate transformations and can be dropped. Substituting the expression for \textit{W}, all three equations derived from Einstein's field equations then assume
the same form, which is
\begin{equation}\label{M1}
\nu'' - \frac{3}{2}\nu'^2+2c_1\nu' - \frac{c_1^2}{2} + \kappa \alpha = 0
\end{equation}
where $c_1$ is a constasnt of integration and $\kappa$ Einstein's constant, with $\alpha$ a constant indicating the strength of the massless
scalar field. Equation \ref{M1} can be solved two ways: (1) by an ansatz, and (2) by completing the square.  The ansatz is given by
guessing that 
\begin{equation}
\nu = f + \beta u
\end{equation}
where \textit{f} is a function of \textit{u} and $\beta$ is a constant.

The general metric derived from an ansatz is given by
\begin{equation} \label{N4}
ds^2 = \frac{e^{\beta u}}{\left( e^{-\gamma u}+ B\right)^{2/3}}dt^2 -
\frac{e^{-\gamma u}}{\left(e^{-\gamma u} + B\right)^2}du^2
-\frac{e^{(-\frac{1}{2}\beta-\frac{1}{2}\gamma) u}}{\left(e^{-\gamma u}+ B\right)^{2/3}}\left(dx^2+dy^2\right)
\end{equation}
where the condition
\begin{equation} \label{M13}
\beta = -\frac{\gamma}{3} \pm \left( \frac{4\gamma^2}{9}- \frac{8}{3} \kappa \alpha\right)^{\frac{1}{2}}
\end{equation}
must be satisfied, along with the conditions $\gamma = 2c_1-3\beta$ and
 $\kappa \alpha = \frac{3}{2}\beta^2 + \frac{1}{2}c_1^2 - 2c_1\beta$, which hold identically.
In addition, the field equations put a condition on $\gamma$:
\begin{equation}\label{cosmo}
-\frac{1}{3} \gamma^2B = \Lambda
\end{equation}
For concreteness, the cosmological constant may be assumed to be carried by the constant $B$.  Hence the metric in Equation \ref{N4} 
is a family of solutions for a minimally-coupled plane-symmetric massless scalar field in general relativity, with cosmological constant.
Here the cosmological solutions are desired, which can be found by analytic continuation.  Following Vaidya-Som \cite{Vaidya}:
the solutions can be transformed using $\beta=i \chi$, $u=iv$, $\gamma=i\delta$, and $t=iw$:
\begin{equation}
ds^2= \frac {e^{\delta v}} {\left(e^{\delta v} + B \right)^{2}} dv^2 - \frac {e^{-\chi v}} 
{\left(e^{\delta v} + B \right)^{2/3}} dw^2 - \frac {e^{\left(\frac{1}{2}\alpha + \frac{1}{2}\delta \right)v}} {\left(e^{\delta v} + B \right)^{2/3}}\left(dx^2+dy^2\right)
\end{equation}

The first term, which now corresponds to the time-time component of the metric, can be transformed by
\begin{equation}
d\tau= \frac {e^{\frac{1}{2}\delta v}} {\left(e^{\delta v} + B \right)} dv 
\end{equation}

There are now three cases to be considered, corresponding to $B=0$, $B>0$, and $B<0$. Graphs of all cosmologies are given in 
Figure 1.
\subsection{Case 1: $B = 0$}
By Equation \ref{cosmo} the cosmological constant is zero and the metric is given by
\begin{equation}ds^2= d\tau^2 - \left[\frac {\delta} {2}\left(\tau-\epsilon\right)\right]^{2/3} \left(dx^2+dy^2+dw^2\right)
\end{equation}
This metric is divergent in its first derivative at $\tau = \epsilon$.  After that time, it expands forever,
although less rapidly than the Robertson-Walker dust or radiation solutions.
\subsection{Case 2: $B >0$}
By Equation \ref{cosmo}, this metric corresponds to a negative cosmological constant.  The metric can be algebraically reduced to the form

\begin{equation}
ds^2= d\tau^2 - \frac{1}{2\sqrt{B}}\sin^{2/3}\left(\delta \sqrt{B}\left(\tau-\epsilon\right)\right)
 \left(dx^2+dy^2+dw^2\right)
 \end{equation}
It's evident that this metric represents a cyclic universe, with repeated epoques of expansion and contraction.

\subsection{Case 3: $B < 0$}
By Equation \ref{cosmo} this metric corresponds to a positive cosmological constant.  It can be algebraically reduced to the form

\begin{equation}
ds^2 = d\tau^2 -  \tanh^{2/3}\left(\delta \sqrt{|B|}(\tau-\epsilon)\right)
\left(dx^2 + dy^2 + dw^2\right)
\end{equation}

This universe can be considered as one that was flat in the infinite past and collapsed, and then expanded again and
evolved into a flat universe. It collapses and reexpands, but only once, as opposed to numerous times for the cyclic universe.
So for the three different choices of cosmological constant, three distinct universes can be obtained: one that expands forever,
one that is cyclic, and finally one that starts flat in the infinite past, collapses, and then evolves
asymptotically to flat space.

\section{Cosmologies from the Second Solution}
A second solution for the metric can be found by completing the square in Equation \ref{M1}. Completing the square would appear
to be the most natural choice, but leads to a solution unrelated to the less general exact solutions. 
Start again from Equation \ref{M1} and complete the square:
\begin{equation}
\nu'' = \frac{3}{2} \left( \nu' - \frac{2}{3}c_1\right)^2 - \frac{1}{6}c_1^2 - \kappa \alpha
\end{equation}
Now define $p' = \nu' - \frac{2}{3}c_1$. The above equation becomes:
\begin{equation} \label{N1}
p'' = \frac{3}{2} p'^2 - |\gamma|
\end{equation}
where $\gamma = \frac{c_1^2}{6}+ \kappa \alpha$. Note here that $\gamma$ is positive definite, because one term
is squared and $\alpha$ must be positive so as to yield a positive energy for the scalar field stress energy. (Relaxing
this condition, which may serve some function in studies of exotic matter or fields, would lead to two other solutions. See
also section 3.3, where negative energy gives rise to inflationary cosmologies.)
 This differential equation is mathematically similar to the problem of an object falling under constant acceleration
 in the presence of air friction. To solve Equation \ref{N1}, substitute
\begin{equation}
p' = \sqrt{\frac{2|\gamma|}{3}} \coth \theta
\end{equation}
This procedure, as outlined in \cite{Vuille}, leads to the metric
Obtain given by
\begin{equation} \label{N6}
 ds^2 = \frac{e^{\frac{2}{3}c_1 u }}{\left|\sinh \left(-\delta u + b \right)\right|^{2/3}}dt^2 -
\frac{du^2}{\left|\sinh \left(-\delta u + b \right)\right|^{2}} -\frac{e^{-\frac{1}{3}c_1 u}}
{\left|\sinh \left(-\delta u + b \right)\right|^{2/3}} \left(dx^2 + dy^2\right)
\end{equation}
where $\delta = \sqrt{3|\gamma|/2}$. Metric
and Riemann tensor components diverge at $u = b/\delta$. As developed in \cite{Vuille} the following condition holds:
\begin{equation}\label{N5}
 \frac{c_1^2}{6} + \kappa \alpha = 2\Lambda 
\end{equation}

From Equation \ref{N5} it is evident only $\Lambda >0$ is permitted, and that as
 $\Lambda \rightarrow 0$, $c_1, \alpha, \gamma \rightarrow 0$, yielding flat space. In view of this last
relationship it appears this solution is special and distinct from the previous class, and doesn't generalize
the solutions of Taub, Novotny and Horsky, and Singh.

Yet another distinct solution can be obtained by making the substitution
\begin{equation}
p'=\sqrt{\frac{2|\gamma|}{3}}\tanh \theta
\end{equation}
Just as in the problem of an object falling under constant acceleration through a gravity field,
this assumption leads to a different solution that is complementary to that obtained previously.

The solution is
\begin{equation} \label{N6}
 ds^2 = \frac{e^{\frac{2}{3}c_1 u }}{\left|\cosh \left(-\delta u + b \right)\right|^{2/3}}dt^2 -
\frac{du^2}{\left|\cosh \left(-\delta u + b \right)\right|^{2}} -\frac{e^{-\frac{1}{3}c_1 u}}
{\left|\cosh \left(-\delta u + b \right)\right|^{2/3}} \left(dx^2 + dy^2\right)
\end{equation}
Unlike the $\sinh$ solution, this solution does not result in divergences in the metric
components when the argument goes to zero.
 
As before, these solutions can be transformed to cosmologies via the complex transformation
as in Vaidya and Som \cite{Vaidya}. Further, to avoid imaginary terms in the metric,
one or more constants would also have to be chosen to be imaginary.  The resulting spacetimes are non-isotropic plane-symmetric
cosmologies. With the $\cosh$ solution given by Equation \ref{N6}, make the analogous Vaidya-Som
coordinate transformations $t \rightarrow iw$, $u \rightarrow iv$, and $\delta \rightarrow i \delta$,
setting $c_1 = 0$ to make the solution isotropic. This process has the consequence of requiring that both $\alpha < 0$ and $\Lambda < 0$.
The metric can then be converted to Robertson-Walker form with 
\[ \frac{dv}{\cosh(\delta v + b)} = d \tau\] which, after some algebra, is given by
\begin{equation}\label{Z}
 ds^2 = d \tau^2 -\cos^{2/3}(\delta (\tau - c_0)) \left(dx^2 + dy^2 + dw^2\right)
\end{equation}
where $c_0$ is an arbitary constant.  Equation \ref{Z} is a cyclic universe similar to the Robertson-Walker dust
cosmology. It is essentially the same universe as described in Case 2, above.

The $\sinh$ solution yields a very different cosmology.  After making the same substitutions as before, convert to
 Robertson-Walker form with
\begin{equation}\label{Z2}
 \frac{-dv}{\sinh(\delta v + b)} = d\tau
\end{equation}
This results in a metric given by
\begin{equation}\label{Z3}
 ds^2 = d \tau^2 -\left(e^{2\delta \tau} - 1\right)^{1/3}\left(dx^2 + dy^2 + dw^2\right)
\end{equation}
The solution of Equation \ref{Z3} exhibits rapid inflation in the early universe, followed by continued exponential
 expansion at all later times.(The sign in Equation \ref{Z2} can be chosen otherwise, and results in the time-reverse of this solution.)
  Figure 2 is a logarithmic graph that illustrates the similarities and differences between the standard Concordance model and this dark energy
spacetime. \cite{Edwards}, \cite{Wright}

\section{Concluding Remarks}
Several solutions to the problem of a plane-symmetric cosmology in general relativity
with massless scalar field and cosmological constant have been found. The first of these solutions involves two parameters, $\beta$
and $\gamma$, which are related to a non-trivial integration constant, designated $c_1$. Despite the fact that the underlying
manifold is plane symmetric, cosmologies analogous to the Robertson-Walker cosmologies for spherical, plane, and hyperbolic symmetry were found.
In addition, a spacetime exhibiting very rapid inflation in the early universe followed by continued exponential
inflation was found. This latter solution expands more rapidly in the early universe than does the Concordance model, but at later
times agrees closely with it. In this solution the cosmological constant and scalar field are mutually dependent, each
vanishing if the other does. Further study of these spacetimes would be of interest, as they appear to have some of the
features required by cosmological observations.

Analogous exact solutions for spherical and hyperbolic symmetry would be desirable.  The current work, for example, bears similarity
 to work done on spherically-symmetric scalar fields by Wyman \cite{Wyman} and others, \cite{fisher}-\cite{Zhang} who
 independently found solutions for spherically-symmetric scalar fields with $\Lambda = 0$. These equations have been examined
 and are at a higher level of complexity. Nonetheless, there is hope for such solutions, and an effort is underway.

\section{Acknowledgements}
The authors are grateful for the comments of the former editor of General Relativity and Gravitation, G. F. R. Ellis, on the 
seed paper from which this current work was derived.

\end{document}